\documentclass[
reprint,
preprintnumbers,
superscriptaddress,
amsmath,amssymb,
aps,physrev,
floatfix,
]{revtex4-2}

\usepackage{graphicx}
\usepackage{dcolumn}
\usepackage{bm}
\usepackage{color}
\usepackage{siunitx}
\usepackage{physics}
\usepackage{here}
\usepackage{xcolor}
\usepackage{etoolbox}

\newcommand{\mrm}[1]{\mathrm{#1}}
\newcommand{\ii}{\mrm{i}}
\newcommand{\ee}{\mrm{e}}

\renewcommand{\today}{July 29, 2026}
\begin{document}

\title{\textbf{Tip-Enhanced Vibrational Ladder Climbing in Surface Molecular System}}

\author{Tatsuto Mochizuki}
\affiliation{Institute for Molecular Science, National Institutes of Natural Sciences, Okazaki, Aichi 444-8585, Japan}
\affiliation{Graduate Institute for Advanced Studies, SOKENDAI, Okazaki, Aichi 444-8585, Japan}
\author{Shota Takahashi}
\affiliation{Institute for Molecular Science, National Institutes of Natural Sciences, Okazaki, Aichi 444-8585, Japan}
\author{Atsunori Sakurai}
\email{asakurai@ims.ac.jp}
\affiliation{Institute for Molecular Science, National Institutes of Natural Sciences, Okazaki, Aichi 444-8585, Japan}
\affiliation{Graduate Institute for Advanced Studies, SOKENDAI, Okazaki, Aichi 444-8585, Japan}
\author{Toshiki Sugimoto}
\email{toshiki-sugimoto@ims.ac.jp}
\affiliation{Institute for Molecular Science, National Institutes of Natural Sciences, Okazaki, Aichi 444-8585, Japan}
\affiliation{Graduate Institute for Advanced Studies, SOKENDAI, Okazaki, Aichi 444-8585, Japan}

\date{\today}

\begin{abstract}
Achieving high-lying vibrational states is essential for actively controlling molecular reactions.
We demonstrate vibrational ladder climbing of CO adsorbed on Pt(111) within the plasmonic tip--substrate nanogap formed in a scanning tunneling microscope, detected via tip-enhanced sum-frequency generation (TE-SFG).
As the infrared pulse energy increases, hot-band peaks appear sequentially up to the 3--4 transition, indicating the stepwise population of higher vibrational states.
Numerical analysis using the optical Bloch equations captures the observed energy dependence of these features.
These results demonstrate the capability of TE-SFG to probe vibrational ladder climbing in surface molecular systems and suggest a promising route toward accessing high-lying vibrational states and controlling vibrational excitation at the nanoscale.
\end{abstract}

\maketitle

The generation of high-lying vibrational states is fundamental to understanding vibrational energy flow and controlling chemical reaction dynamics in the electronic ground state. Driving a specific intramolecular vibration to large amplitudes along the reaction coordinate may enable reaction control that selectively cleaves specific chemical bonds (i.e., mode-selective chemistry) \cite{JPC1996_Crim,ARPC2016_Liu} and, more generally, control of vibrational quantum states \cite{NJOP2010_Brif}.
However, because bond dissociation energies are typically much larger than the energy of a single vibrational quantum, multiple vibrational excitations are required to actively control molecular reactions. Furthermore, vibrational energy relaxes on the picosecond timescale in condensed phases, preventing its efficient accumulation.
To address these difficulties, sequential multi-quantum vibrational excitation, referred to as ``vibrational ladder-climbing'' (VLC), is essential for populating high-lying vibrational states. Advances in infrared (IR) laser technology have enabled VLC and subsequent bond cleavage in gas phases \cite{CPL2002_Windhorn} and have extended VLC to condensed phases \cite{CPL2004_Witte,PNAS2004_Ventalon,PRL2010_Jewariya,JPCL2016_Kemlin}. Furthermore, IR pulse-shaping techniques have been used to demonstrate coherent control of high-lying vibrational states \cite{PRL2007_Strasfeld}. Despite these advances, the low absorption cross-section of molecular vibrations limits such approaches to only a few molecular systems.

Optical field enhancement due to plasmonic resonance offers a promising strategy to overcome this limitation. Specifically, periodic metallic nanostructures fabricated by electron beam lithography effectively enhance weak vibrational excitation through plasmonic near fields, as the resonance frequency can be tuned by adjusting the rod length \cite{CR2017_Neubrech}. While plasmon-enhanced nonlinear vibrational spectroscopy has been demonstrated \cite{CR2017_Neubrech,PRL2015_Selig,JPCC2017_Morichika,JPCC2018_Mackin,JCP2020_Chuntonov}, extended to VLC in surface molecular systems \cite{PCCP2016_Kraack}, and applied to molecular dissociation \cite{NC2019_Morichika}, near-field enhancement in such periodic nanostructures is typically confined to a limited spectral range, and enhancement efficiency drops rapidly when the wavelength is detuned from the resonance frequency. Because molecular vibrations are inherently anharmonic, transition frequencies between adjacent states redshift as the vibrational quantum number increases, suggesting that field enhancement over a broader spectral range is beneficial for VLC, especially in molecular systems with significant anharmonicity.

Another approach to enhancing the IR excitation field is nanofocusing along the tapered shaft of a metallic tip (the antenna effect) \cite{PRL2004_Stockman}. Our previous work showed that incident IR fields are enhanced over a broad spectral range as surface plasmon polaritons (SPPs) propagate along the tapered shaft toward the tip apex, with more pronounced enhancement at longer wavelengths \cite{JPCL2023_Takahashi}. Additionally, when vibrational responses are upconverted into visible signals via sum-frequency generation (SFG) with a visible or near-IR pulse \cite{JCP2025_Shen}, the SFG emission is enhanced by the gap-mode plasmon resonance formed between the metallic tip and the substrate \cite{JPCL2023_Takahashi,NL2025_Sakurai}.

Here, we report the observation of VLC of surface molecules within the nanogap between a scanning tunneling microscope (STM) tip and a metallic substrate using tip-enhanced SFG (TE-SFG) \cite{NL2025_Sakurai,JCP2026_Sakurai,JPCC2026_Takahashi}. Unlike conventional IR pump--probe spectroscopy, which detects small absorption changes induced by pump excitation, TE-SFG provides background-suppressed detection by upconverting vibrational responses into photons at different frequencies. This combination of high sensitivity and nanoscale field confinement circumvents spatial averaging over a focal area \cite{JPCC2026_Takahashi}, enabling VLC beyond the diffraction limit. As IR pulse energy increases, hot-band peaks emerge sequentially up to the 3--4 transition, demonstrating stepwise population transfer to higher-lying vibrational states. These results show that tip-enhanced near fields can drive VLC in surface molecular systems and that TE-SFG provides a unified nanoscale platform for both the excitation and probing of high-lying vibrational states.

\begin{figure}
    \centering
    \includegraphics[width=\linewidth]{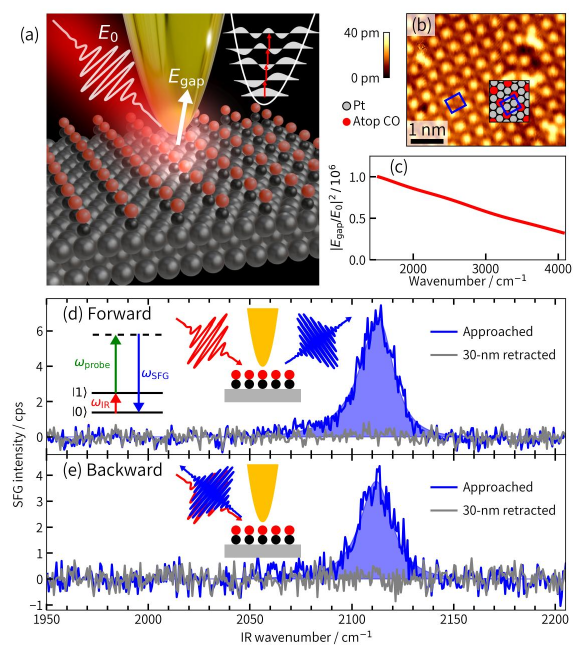}
    \caption{(a) Schematic of tip-enhanced VLC. CO on Pt(111) is excited by an IR near field in the tip--substrate nanogap with an electric field component normal to the substrate surface ($\bm{E}_\mathrm{gap}$).
    (b) STM image of CO/Pt(111) at \SI{0.5}{ML} coverage
        acquired
        in constant-current mode
        at a sample bias voltage of $\SI{0.25}{V}$ and
        a tunneling current of $\SI{0.5}{nA}$.
        The image was preprocessed with the ``Stripe Removal'' program \cite{OE2025_Rottmayer}.
        The inset is a schematic representation of Pt atoms and CO molecules.
        The blue rectangles indicate the unit cell of the c($\sqrt{3}\times2$)rect structure.
    (c) Enhancement factor as a function of the incident wavenumber calculated by electric field simulations (see the main text for details).
    (d) Forward- and (e) backward-scattered TE-SFG spectra.
    SFG intensities are normalized to counts per second (cps).
    The blue curves represent the spectra taken with the substrate positioned close to the tip apex at a sample bias of \SI{0.25}{V}, a tunneling current of \SI{0.5}{nA}, and an IR pulse energy of \SI{0.4}{pJ}.
    The gray curves represent the spectra taken with the substrate retracted by \SI{30}{nm} from the tip apex and an IR pulse energy of \SI{60}{pJ}.
    The probe pulse energy remained constant at \SI{14}{pJ} for both cases.
    The fitted curves are superimposed as blue filled curves.
    The schematic diagram in (d) illustrates the SFG process, which  is described as vibrational excitation by the IR pulse followed by an anti-Stokes Raman scattering process.}
    \label{fig:1}
\end{figure}

Experiments were conducted using our TE-SFG nanoscopy system combined with STM \cite{JPCL2023_Takahashi, NL2025_Sakurai, JCP2026_Sakurai, JPCC2026_Takahashi, NC2026_Takahashi}.
As a model system for demonstrating tip-enhanced VLC, we employed carbon monoxide (CO) molecules adsorbed on an atomically flat Pt(111) surface (Fig.~\ref{fig:1}(a)). We prepared a clean Pt(111) surface through repeated cycles of Ar sputtering, O\textsubscript{2} annealing, and flashing to \SI{1000}{K}. Formation of the well-known \SI{0.5}{ML} CO coverage was confirmed by observation of the c($\sqrt{3}\times2$)rect structure in an STM image (Fig.~\ref{fig:1}(b)), consistent with previous reports \cite{CPL1996_Song, JPCC2013_Yang}. The corresponding far-field (FF) SFG spectrum without tip enhancement exhibits characteristic features of \SI{0.5}{ML} CO/Pt(111) (see Supplemental Material Sec.~S4 \cite{SM}\nocite{AANM2021_Cui, PR1929_Morse, JMS1975_Mantz, JCP1983_Vasan}). To create a nanogap that enables strong near-field enhancement, a smooth and sharpened plasmonic Au tip \cite{JPCC2018_Yang} (see Supplemental Material Sec.~S2 \cite{SM}) was mounted in a low-temperature STM system (USM1400, Unisoku) operated at liquid-nitrogen temperature ($\mathord{\sim}\SI{80}{K}$) under ultrahigh vacuum ($\mathord{\sim}\SI{e-8}{\pascal}$). The resulting tip--substrate nanogap achieved an estimated intensity enhancement exceeding $10^5$ across the broad IR spectral range due to nanofocusing along the tip shaft (Fig.~\ref{fig:1}(c)) \cite{JPCL2023_Takahashi,NL2025_Sakurai,JPCC2026_Takahashi}. This broadband feature provides a significant advantage for accessing high-lying vibrational states through the VLC process compared to periodic metallic nanostructures (see Supplemental Material Sec.~S1 \cite{SM}). The IR pulse, with a central wavenumber of approximately \SI{2100}{cm^{-1}}, resonantly excites the atop CO stretching mode, and a narrowband 1030-nm pulse with a bandwidth of approximately \SI{6}{cm^{-1}} upconverts the resulting vibrational response into SFG emission. Hereafter, we refer to this 1030-nm upconversion pulse as the ``probe'' pulse. Both pulse energies were adjusted to prevent optical damage to the tip (IR: $<$\SI{40}{pJ}, probe: $<$\SI{20}{pJ}). Further details of the optical setup are provided in Supplemental Material Sec.~S3 \cite{SM}.

\begin{figure}
    \centering
    \includegraphics[width=\linewidth]{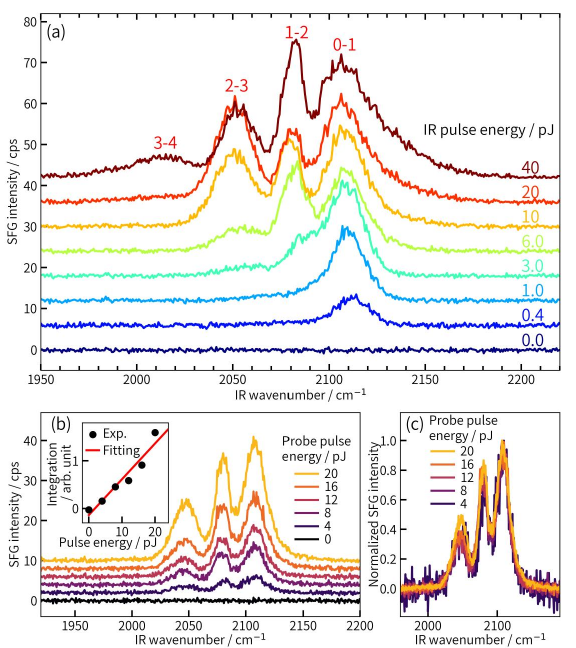}
    \caption{(a) IR pulse energy dependence of forward-scattered TE-SFG spectra for CO/Pt(111). The probe pulse energy was maintained at \SI{14}{pJ}. Backward-scattered spectra are shown in Supplemental Material Sec.~S5 \cite{SM}.
    (b, c) Probe pulse energy dependence of TE-SFG spectra with IR pulse energy fixed at \SI{10}{pJ}. Signal intensities are normalized to the peak intensity (c). The inset in (b) shows the integrated intensity of TE-SFG spectra as a function of the probe pulse energy. The red line is a linear fit to the data.}
    \label{fig:2}
\end{figure}

Figure~\ref{fig:1}(d) illustrates a typical SFG spectrum obtained in the forward-scattering direction. When the tip--substrate distance was \SI{30}{nm}, no appreciable signal was detected using an IR pulse energy of \SI{60}{pJ} (gray curve). However, when the tip--substrate distance was reduced to less than \SI{1}{nm}, an SFG signal appeared as a single resonant peak near \SI{2110}{cm^{-1}} even with an IR pulse energy of \SI{0.4}{pJ} (blue curve). The observed peak corresponds to the 0--1 transition of the CO stretching mode. A similar spectrum was also recorded in the backward-scattering direction, where phase matching is not satisfied (Fig.~\ref{fig:1}(e), blue curve). These results indicate that the measured signal originates from a near-field effect rather than a far-field contribution \cite{JPCL2023_Takahashi,NL2025_Sakurai,JPCC2026_Takahashi,NC2026_Takahashi,JCP2026_Sakurai}.

We measured TE-SFG spectra at increasing IR pulse energies (Fig.~\ref{fig:2}(a)). As the IR pulse energy surpassed \SI{3}{pJ}, a new peak appeared at \SI{2080}{cm^{-1}} on the low-wavenumber side of the 0--1 band. This peak is attributed to the 1--2 vibrational transition, which is red-shifted from the 0--1 resonance due to vibrational anharmonicity. Furthermore, additional peaks appeared with increasing IR pulse energy, corresponding to the 2--3 transition at \SI{2050}{cm^{-1}} and the 3--4 transition at \SI{2010}{cm^{-1}}. These multi-peak structures were not observed in the FF-SFG measurements under the present experimental conditions, despite the much stronger incident IR pulse energy of \SI{1}{nJ} (see Supplemental Material Sec.~S4 \cite{SM}). With the IR pulse energy fixed, the signal intensity increased with increasing probe pulse energy (Fig.~\ref{fig:2}(b)), while the spectral shape remained unchanged (Fig.~\ref{fig:2}(c)). The detection of the 3--4 transition indicates that the third excited vibrational state is populated prior to the probe pulse upconversion. The observed hot bands cannot be explained by a quasi-equilibrium thermal population. Even at the desorption temperature of CO from the Pt(111) surface ($\mathord{\sim}\SI{400}{K}$ \cite{SS1977_Ertl}), the Boltzmann factor for the third vibrational state (\SI{8240}{cm^{-1}} above the ground state) amounts to only $\mathord{\sim}10^{-13}$. Together with the probe-energy independence of the spectral shape (Fig.~\ref{fig:2}(b,c)), these observations indicate that the hot bands originate from non-thermal, IR-driven stepwise excitation. The hot bands almost completely disappear at an IR--probe delay of \SI{5}{ps}, which is far shorter than the \SI{20}{ns} interval between successive laser pulses at the \SI{50}{MHz} repetition rate. This result rules out population accumulation over successive pulses and supports vibrational ladder climbing driven by a single IR pulse (see Supplemental Material Sec.~S8 \cite{SM}). These results demonstrate that the sequential vibrational transitions were induced within the tip--substrate nanogap by near-field enhancement of the IR excitation, thereby providing the first observation of tip-enhanced VLC.

\begin{figure}
    \centering
    \includegraphics[width=\linewidth]{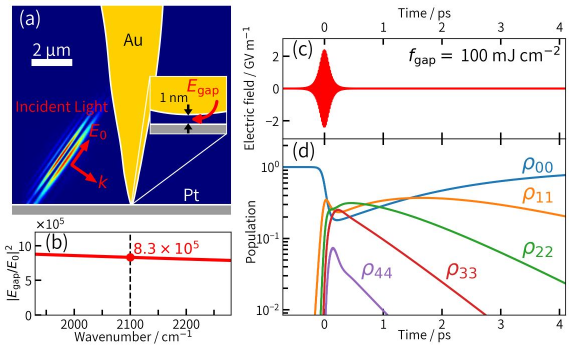}
    \caption{(a) FDTD simulation setup for calculating the electric field enhancement in the tip--substrate nanogap.
             (b) Intensity enhancement factor ($|E_{\text{gap}}/E_0|^2$) at the center of the nanogap as a function of the incident light wavenumber.
                 The intensity enhancement factor is estimated to be $\num{8.3e5}$ near the CO vibrational resonance.
             (c) Temporal profile of the electric field in the nanogap under irradiation by an incident pulse with fluence $f_{\mathrm{gap}}=\SI{100}{mJ.cm^{-2}}$.
             (d) Time evolution of populations of the $v=0-4$ states calculated using the multilevel optical Bloch equations when $f_{\text{gap}}=\SI{100}{mJ.cm^{-2}}$.
             }
    \label{fig:3}
\end{figure}

Next, we estimate the electric field enhancement in the tip--substrate nanogap by performing finite-difference time-domain (FDTD) simulations. The simulated system comprises a Au tip positioned 1 nm above a Pt substrate, representing the nanogap structure used in our experiments (Fig.~\ref{fig:3}(a)). The refractive indices of Au and Pt were taken from previously reported values \cite{H1985_Palik,PRB2012_Olmon}, and the tip geometry was modeled from scanning electron microscopy (SEM) images acquired after the TE-SFG measurements (see Fig.~S2 in Supplemental Material for SEM images of the tip). Illuminating the nanogap with \textit{p}-polarized light ($E_0$) at an incident angle of \SI{55}{\degree} and monitoring the enhanced near field within the nanogap ($E_\mathrm{gap}$) yields an intensity enhancement factor of $\left|E_\mathrm{gap}/E_0\right|^2 \approx 10^6$ near the CO vibrational resonance (Fig.~\ref{fig:3}(b)). Under this strong near-field enhancement, the near-field fluence within the nanogap is estimated to reach approximately \SI{100}{mJ.cm^{-2}} for an IR pulse energy of \SI{40}{pJ}. This value matches or exceeds those used in prior VLC studies of CO adsorbed on metal surfaces \cite{CPL2000_Hess,SS2005_Zhang,JPCC2025_Omiya}.

From the deduced gap fluence of \SI{100}{mJ.cm^{-2}} and the IR pulse duration (FWHM) of \SI{120}{fs}, the peak amplitude of the IR near field within the nanogap ($E_{\text{IR}}\left(t\right)$) is estimated at \SI{2.4}{GV/m} (Fig.~\ref{fig:3}(c)). Using this field as the excitation source, we performed numerical simulations of multiphoton vibrational excitation dynamics by solving the multilevel optical Bloch equations \cite{PRL2010_Jewariya, NC2019_Morichika}.
\begin{align}
    \pdv{\rho_{vw}(t)}{t}&=-\frac{\ii}{\hbar}\left[H_0-\mu E_{\text{IR}}(t), \rho(t) \right]_{vw}\notag \\
                      &\quad-\frac{\Gamma_v+\Gamma_w}{2}\rho_{vw}(t) +\delta_{vw}\Gamma_{v+1}\rho_{v+1,v+1}(t),
\end{align}
where $H_0$ is the vibrational Hamiltonian, $\mu$ is the dipole moment, $\Gamma_v$ is the population relaxation rate from level $v$ to $v-1$, and $\delta_{vw}$ is the Kronecker delta. Since the observed vibrational progression exhibits nearly uniform spacing between adjacent peaks (Fig.~\ref{fig:2}), the vibrational potential of the CO stretching mode is well modeled by a Morse potential, where the vibrational eigenenergy of the state $\ket{v}$ is given by $E_v=\hbar\omega_{\text{e}}\left(v+1/2\right)-\hbar\omega_{\text{e}}\chi\left(v+1/2\right)^2$. The equilibrium frequency $\omega_{\text{e}}$ and the anharmonicity constant $\chi$ were obtained by linear fitting of the peak positions as a function of the vibrational quantum number: $\omega_{\text{e}}/(2\pi c)=\SI{2144.8}{cm^{-1}}$ and $\chi=\num{7.6e-3}$ (see Supplemental Material Sec.~S6 \cite{SM}). For the transition dipole moments between states $\ket{v}$ and $\ket{v+1}$, we used the relation $\mu_{v+1,v}=\bra{v+1}\mu\ket{v}=\mu_{10}\bra{v+1}\xi\ket{v}/\bra{1}\xi\ket{0}$, where $\xi$ represents the vibrational normal coordinate, and assumed the relation $\mu\propto\xi$. The matrix elements $\bra{v+1}\xi\ket{v}$ were calculated using vibrational wavefunctions of a Morse potential (see Supplemental Material Sec.~S6 \cite{SM}), and the value of $\mu_{10}$ ($=\SI{0.2}{D}$) was adopted from the literature \cite{JCP1991_Beckerle}. The decay rate of the state $\ket{v+1}$ was assumed to be given by $\Gamma_{v+1}=\Gamma_1(\bra{v+1}\xi\ket{v})^2/(\bra{1}\xi\ket{0})^2$ \cite{JCP1998_Jakob}, and $\Gamma_1$ was set to \SI{0.5}{ps^{-1}} based on a previous study \cite{JCP1991_Beckerle}. The time evolution of the density matrix, including vibrational levels up to $v=6$, was numerically calculated using a fourth-order Runge--Kutta method.

\begin{figure}
\centering
\includegraphics[width=\linewidth]{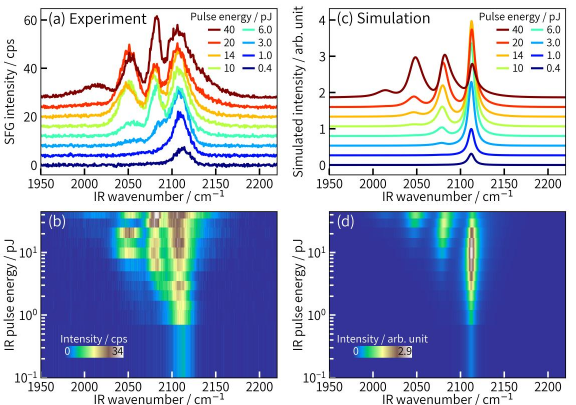}
    \caption{Experimental and simulated TE-SFG spectra of CO on Pt(111).
    (a) Experimental TE-SFG spectra at various IR pulse energies.
    (b) 2D map of experimental TE-SFG spectra as a function of wavenumber and IR pulse energy.
    (c) Simulated TE-SFG spectra at various IR pulse energies.
    (d) 2D map of simulated TE-SFG spectra as a function of wavenumber and IR pulse energy.
    The spectra in panels (a) and (c) are vertically offset for clarity.}
    \label{fig:4}
\end{figure}

To simulate the SFG spectra, we considered a process in which the vibrational response induced by the IR pulse interacts with the subsequent probe field $E_{\text{probe}}(t)$ to generate SFG emission through anti-Stokes Raman scattering. The polarization responsible for the SFG radiation is given by
\begin{equation}
    P(t) = \mathrm{Tr}\left[\alpha \rho(t)\right] E_\mathrm{probe}(t),
\end{equation}
where $\alpha$ is the polarizability. Assuming electronic non-resonance in Raman transitions, the Raman polarizabilities are given by $\alpha_{v+1,v}=\bra{v+1}\alpha\ket{v}=\alpha_{10}\bra{v+1}\xi\ket{v}/\bra{1}\xi\ket{0}$ \cite{PRL2024_Batignani}. Using the convention $E(t)\propto\ee^{-\ii\omega t}$ \cite{Hamm_Zanni_2011,JCP2026_Sakurai}, we define the frequency-domain SFG field as
\begin{equation}
    \tilde{E}_\mathrm{SFG}(\omega) \propto \int_{-\infty}^\infty \dd t\, P(t)\,\ee^{\ii\omega t}.
\end{equation}
The simulated SFG intensity spectrum corresponding to the experiments is given by  $|\tilde{E}_{\text{SFG}}(\omega)|^2$.

Figure~\ref{fig:3}(d) depicts the time evolution of the populations of the five lowest vibrational states obtained by solving the optical Bloch equations (Eq.~(1)). Following IR pulse irradiation, the ground-state population $\rho_{00}$ decreases, while excited-state populations grow. We found that higher vibrational states up to $v=4$ are populated at the present near-field fluence ($\SI{100}{mJ.cm^{-2}}$) (Fig.~\ref{fig:3}(d)), supporting the experimentally observed 3--4 transition (Fig.~\ref{fig:2}(a)). 
Simulated SFG spectra calculated at varying IR pulse energies exhibit sequential multi-peak emergence (Fig.~\ref{fig:4}) and pulse-energy-dependent broadening of the 0--1 band (see Supplemental Material Sec.~S7 \cite{SM}), capturing the essential features of the tip-enhanced VLC spectra.

However, there are some discrepancies between the experimental and simulated spectral shapes, particularly in the relative peak amplitudes and linewidths (Fig.~\ref{fig:4}). We suggest that these discrepancies arise from physical effects omitted from our simplified model (Eq.~(1)), including intermolecular dipole--dipole coupling and the resulting delocalized vibrational excitations in the densely packed CO adlayer \cite{Hamm_Zanni_2011}, instantaneous frequency shifts and associated pure dephasing induced by anharmonic coupling between the CO stretching mode and low-frequency modes \cite{JCP2012_Inoue,PRL2016_Inoue}, and the spatial nonuniformity of the tip-enhanced near field. Elucidating these microscopic mechanisms will require further investigation.

Finally, we remark that the highest-lying vibrational transition observed in the present study is the 3--4 transition; however, this limit is imposed by the spectral bandwidth of the IR pulses used in the experiments ($\mathord{\sim}\SI{90}{cm^{-1}}$ FWHM). Indeed, simulations based on the optical Bloch equations predict that the 7--8 transition would be observable if IR pulses with an FWHM bandwidth of \SI{260}{cm^{-1}} were used (see Supplemental Material Sec.~S9 \cite{SM}). In addition, negatively chirped IR pulses can further enhance the efficiency of VLC in an anharmonic vibrational potential \cite{PRL2007_Strasfeld,NC2019_Morichika}.

In summary, by exploiting the antenna effect of the tip to achieve an electric-field intensity enhancement of approximately $10^6$ together with high-sensitivity SFG detection, we observed vibrational ladder climbing in CO/Pt(111) up to the 3--4 transition via TE-SFG. This approach provides a means of probing ultrafast dynamics involving high-lying vibrational states in surface molecular systems and offers a route toward mode-selective chemistry by selectively exciting specific vibrational modes. Furthermore, combining pulse shaping \cite{NC2019_Morichika, PRL2007_Strasfeld, NJOP2010_Brif} may enable more efficient ladder climbing and coherent vibrational control at the nanoscale.

{\it{}Acknowledgments} --
We thank M. Aoyama, T. Kondo, N. Mizutani, T. Kikuchi, and T. Toyoda at the Equipment Development Center, Institute for Molecular Science (IMS), and E. Nakamura at the UVSOR synchrotron facility of IMS for their technical assistance. SEM observation of tips was conducted at IMS,  supported by ``Advanced Research Infrastructure for Materials and Nanotechnology in Japan (ARIM)'' of the Ministry of Education, Culture,  Sports, Science and Technology (MEXT), Proposal Number JPMXP1226MS5014. T.S. acknowledges financial support from JSPS KAKENHI Grant-in-Aid for Scientific Research (S) (26K21748), for Scientific Research (A) (19H00865), and for Transformative Research Areas (A) (24H02205); JST PRESTO (JPMJPR1907); JST CREST (JPMJCR22L2); the grant of OML Project by the National Institutes of Natural Sciences (NINS program No. OML032501); and Special Project by IMS (IMS program 25IMS1101). A.S. acknowledges financial support from JSPS KAKENHI Grant-in-Aid for Scientific Research (B) (23K26548) and for Challenging Research (Exploratory) (26K22784). S.T. acknowledges financial support from JSPS KAKENHI Grant-in-Aid for JSPS Fellows (22KJ3099) and for Early-Career Scientists (26K17768).

{\it{}Data Availability} -- The data are available from the authors upon reasonable request.

\begingroup
\makeatletter
\let\addcontentsline\@gobblethree
\makeatother

\makeatother
\endgroup

\clearpage
\onecolumngrid
\setcounter{page}{1}
\setcounter{section}{0}
\setcounter{subsection}{0}
\setcounter{subsubsection}{0}
\setcounter{figure}{0}
\setcounter{table}{0}
\setcounter{equation}{0}

\renewcommand{\thepage}{S\arabic{page}}
\renewcommand{\thesection}{S\arabic{section}}
\renewcommand{\thesubsection}{\thesection.\arabic{subsection}}
\renewcommand{\thesubsubsection}{\thesubsection.\arabic{subsubsection}}
\renewcommand{\thefigure}{S\arabic{figure}}
\renewcommand{\thetable}{S\arabic{table}}
\renewcommand{\theequation}{S\arabic{equation}}
\renewcommand{\figurename}{Fig.}
\renewcommand{\tablename}{Tab.}

\makeatletter
\def\@hangfrom@section#1#2#3{\@hangfrom{#1#2}#3}%
\def\@hangfroms@section#1#2{#1#2}%
\renewcommand{\bibnumfmt}[1]{[S#1]}
\renewcommand{\citenumfont}[1]{S#1}
\makeatother

\raggedbottom

\begin{center}
{\large\bfseries Supplemental Material for\\
``Tip-Enhanced Vibrational Ladder Climbing in Surface Molecular System''\par}
\vspace{1.2em}
Tatsuto Mochizuki,$^{1,2}$ Shota Takahashi,$^{1}$ Atsunori Sakurai,$^{1,2,*}$
and Toshiki Sugimoto$^{1,2,\dagger}$\\[0.7em]
{\small
$^{1}$Institute for Molecular Science, National Institutes of Natural Sciences,
Okazaki, Aichi 444-8585, Japan\\
$^{2}$Graduate Institute for Advanced Studies, SOKENDAI,
Okazaki, Aichi 444-8585, Japan\\[0.5em]
$^{*}$asakurai@ims.ac.jp; $^{\dagger}$toshiki-sugimoto@ims.ac.jp\\[0.5em]
July 29, 2026}
\end{center}
\vspace{1em}

\tableofcontents
\clearpage

\section{Broadband field enhancement in a tip--substrate system}
\begin{figure}[H]
    \centering
        \centering
        \includegraphics[width=0.8\linewidth]{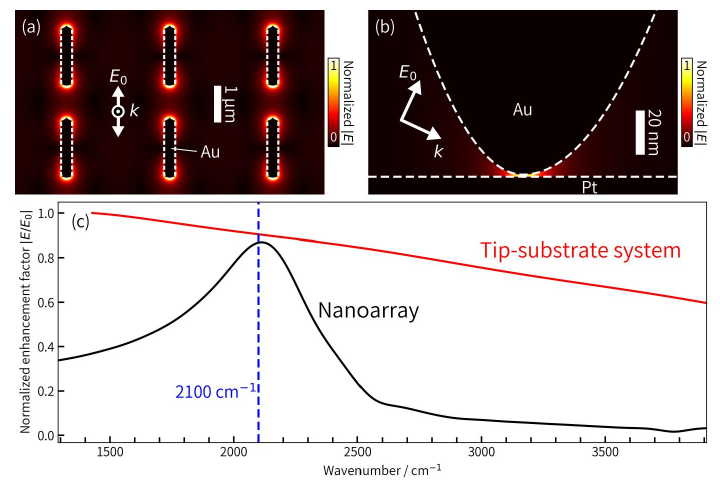}
        \caption{Spatial distribution of the normalized electric field $|E|$ at $\omega/(2\pi c)=\SI{2100}{cm^{-1}}$ in (a) the Au nanoarray and (b) the tip--substrate system.
                 (c) Normalized electric-field enhancement spectra of the tip--substrate gap (red curve) and the nanoarrays (black curve). The enhancement in the tip--substrate system exhibits a broader spectral feature across the IR region compared to that of the nanoarrays.}
        \label{fig:S1}
\end{figure}

In this section, we examine the spectral characteristics of field enhancement, focusing on a comparison between the tip--substrate system and metallic nanoarrays. Electromagnetic field simulations based on the finite-difference time-domain (FDTD) method were performed for both structures (Fig.~\ref{fig:S1}(a), (b)). The metallic nanoarrays were designed based on the structure reported in Ref.~\cite{SI:NC2019_Morichika}: a periodic array of Au nanorods with a length of \SI{1.65}{\micro m}, width of \SI{0.3}{\micro m}, and height of \SI{0.1}{\micro m} was placed on a CaF$_2$ substrate (Fig.~\ref{fig:S1}(a)). The refractive index of the substrate was fixed at \num{1.41}. The array period was set to \SI{2.5}{\micro m} in the longitudinal direction and \SI{2.8}{\micro m} in the transverse direction, and periodic boundary conditions were applied in both directions. Incident light was introduced from the CaF$_2$ substrate side at normal incidence, with the polarization aligned along the long axis of the nanorods. Details of the simulation setup for the tip--substrate system (Fig.~\ref{fig:S1}(b)) are provided in the main text.

The field enhancement spectrum of the nanoarrays exhibits a resonance centered at approximately \SI{2100}{cm^{-1}} (black curve in Fig.~\ref{fig:S1}(c)). In contrast, the tip--substrate nanogap provides broadband field enhancement across the infrared (IR) region (red curve in Fig.~\ref{fig:S1}(c)). This broadband IR response is advantageous for accessing high-lying vibrational states via vibrational ladder climbing (VLC)
because vibrational transitions between adjacent levels shift to lower frequencies as the vibrational quantum number increases. Moreover, the broad enhancement profile of the tip--substrate system may enable VLC for a variety of molecular species on a single platform.

In addition to its broad spectral response, the tip--substrate system offers a clear advantage in spatial selectivity. In the nanoarray structure shown in Fig.~\ref{fig:S1}(a), field enhancement mainly occurs within sub-micron regions at both ends of the longitudinal axis of each nanorod, forming multiple hot spots. In contrast, the tip--substrate nanogap strongly confines the field to a single hot spot, enabling highly localized excitation (Fig.~\ref{fig:S1}(b)). Such localization in the tip--substrate geometry is better suited to driving and probing vibrational excitation at the nanoscale. These results highlight the tip--substrate gap as a promising platform for efficient and spatially selective VLC.

\section{Au tip used in this study}
\begin{figure}[H]
    \centering
        \includegraphics[width=0.9\linewidth]{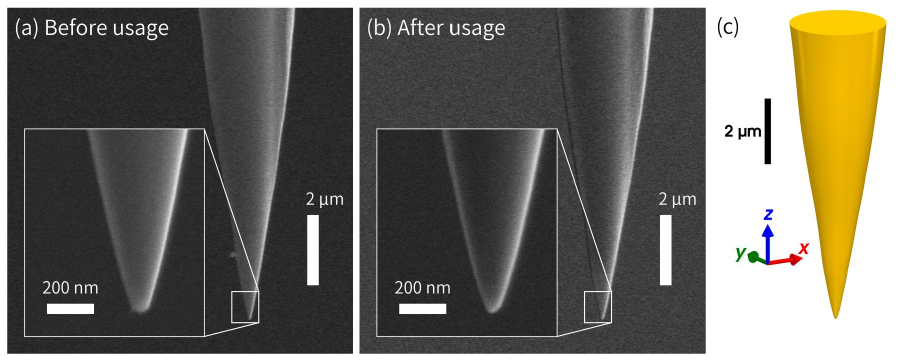}
        \caption{SEM images of the Au tip used in this study: (a) before and (b) after the TE-SFG measurements. These images were obtained using an SU6600 (Hitachi) with an accelerating voltage of \SI{5}{kV} and a working distance of $\mathord{\sim}\SI{10}{mm}$. (c) A three-dimensional model of the tip geometry reconstructed from the post-measurement SEM image in (b).}
        \label{fig:S2}
\end{figure}

The Au tip was fabricated from Au wire by electrochemical etching.
Scanning electron microscope (SEM) images of the Au tip used in this study were acquired before and after the tip-enhanced sum-frequency generation (TE-SFG) measurements (Figs.~\ref{fig:S2}(a) and (b), respectively). No significant changes in the tip geometry were observed, indicating that laser-induced modification of the tip is negligible. A three-dimensional geometric model of the tip (Fig.~\ref{fig:S2}(c)), used in the near-field simulations presented in the main text (Fig.~3(a)), was constructed by digitizing the outline of the post-measurement tip image (Fig.~\ref{fig:S2}(b)).

\newpage
\section{Optical setup for TE-SFG measurements}
\begin{figure}[H]
    \centering
        \includegraphics[width=0.65\linewidth]{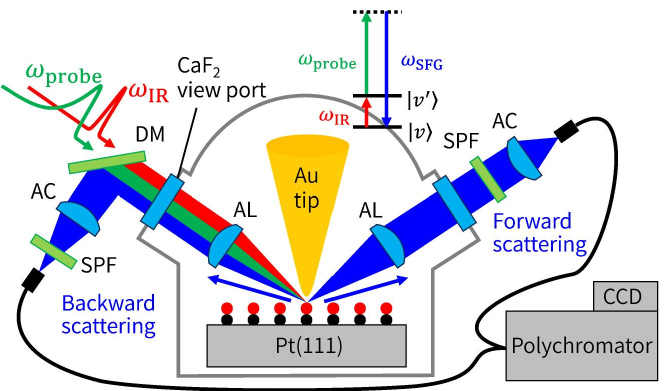}
        \caption{Schematic diagram of the TE-SFG setup.
    The abbreviations are as follows: DM: dichroic mirror, AL: CaF$_2$ aspheric lens, AC: achromatic lens, SPF: short pass filter.}
        \label{fig:S3}
\end{figure}

Details of the optical setup are described elsewhere \cite{SI:JCP2026_Sakurai}, but the pump laser was replaced with Flint (Light Conversion) instead of Monaco (Coherent); here we summarize the essential points. Figure~\ref{fig:S3} illustrates the experimental setup for TE-SFG measurements. The IR ($\omega_\mathrm{IR}$) and 1030-nm upconversion probe ($\omega_\mathrm{probe}$) pulses used for TE-SFG were derived from a Yb:KGW laser (\SI{1030}{nm}, \SI{110}{fs}, \SI{7}{W}, \SI{50}{MHz}; Flint, Light Conversion). The output of this laser was split into two arms using a beam splitter. The first arm was directed into a synchronously pumped optical parametric oscillator (Levante IR, APE) to generate signal (1.3--\SI{2}{\micro m}) and idler (2.1--\SI{5}{\micro m}) pulses. The idler pulse, with a central wavenumber of approximately \SI{2100}{cm^{-1}}, was used to resonantly excite the atop CO stretching mode on a Pt(111) surface. The idler pulse energy was controlled using a pair of linear polarizers (WP25M-IRA, Thorlabs). The second arm from the Yb:KGW laser was passed through an air-spaced Fabry-P\'erot etalon (\#A20, SLS Optics) to narrow the spectral width to $\mathord{\sim}\,\SI{6}{cm^{-1}}$ (full width at half maximum, FWHM), and was used as the upconversion pulse. The IR and upconversion beams were combined collinearly using a dichroic mirror and focused onto the tip--substrate nanogap with a CaF$_2$ aspheric lens at an angle of incidence $\theta_{\text{in}}=\SI{55}{\degree}$. Both incident beams were \textit{p}-polarized. The SFG signal emitted from the nanogap was collected in two directions: forward scattering (specular reflection direction) and backward scattering. These two signals were separately coupled into a bifurcated fiber bundle without polarization selection and directed into a spectrometer (Kymera 328i, Andor). Both signals were focused onto distinct vertical positions on the same electronically cooled CCD detector (iDus 416, Andor), enabling simultaneous measurement.

\section{Far-field SFG spectra of CO/Pt(111)}
\begin{figure}[H]
    \centering
        \includegraphics[width=0.8\linewidth]{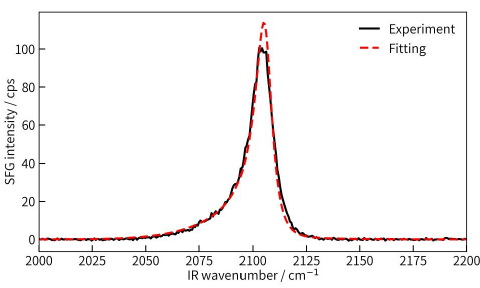}
        \caption{Experimental FF-SFG spectrum of the atop CO stretching mode of \SI{0.5}{ML} CO/Pt(111) obtained at a substrate temperature of $\mathord{\sim}\,\SI{80}{K}$ (black solid curve).
        The red dashed curve shows the result of the curve-fitting analysis.}
        \label{fig:S4}
\end{figure}

Far-field SFG (FF-SFG) measurements for CO/Pt(111) were performed by illuminating the surface while maintaining the tip--substrate distance at \SI{30}{nm} to suppress near-field enhancement. Figure~\ref{fig:S4} shows the FF-SFG spectrum of the atop CO stretching mode obtained with IR and probe pulse energies of \SI{1}{nJ} and \SI{0.4}{nJ}, respectively. For the fitting analysis, we assumed that the vibrationally resonant molecular response is described by a single Lorentzian function, and that the overall FF-SFG spectrum is obtained by the convolution of the molecular vibrational response with the probe pulse spectrum. The peak position and linewidth were estimated to be \SI{2106}{cm^{-1}} and \SI{1.86}{cm^{-1}}, respectively, consistent with previously reported values for \SI{0.5}{ML} CO/Pt(111) at \SI{150}{K}~\cite{SI:JCP1991_Beckerle}.

\begin{figure}[H]
    \centering
        \includegraphics[width=\linewidth]{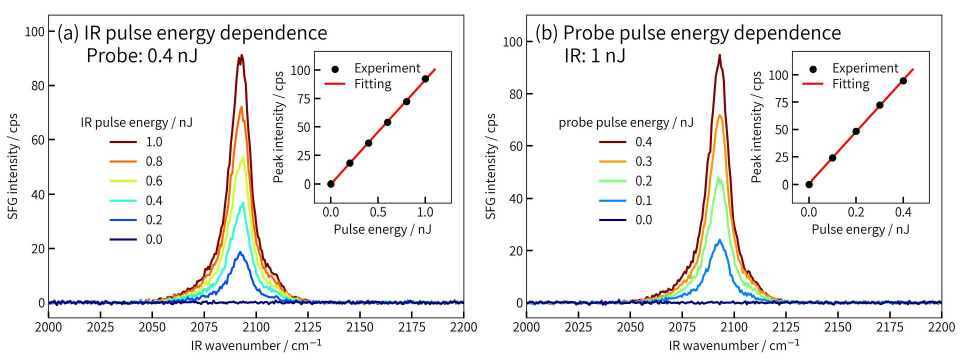}
        \caption{IR (a) and probe (b) pulse energy dependences of FF-SFG signals from \SI{0.15}{ML} CO/Pt(111). The insets in (a) and (b) show the integrated intensities of the FF-SFG spectra plotted as a function of pulse energy (black filled circles), along with their linear fits (red lines). Note that the CO coverage in these FF-SFG experiments (\SI{0.15}{ML}) is lower than the saturation value (\SI{0.5}{ML}), resulting in a slight redshift of the spectral peak compared to that of \SI{0.5}{ML} CO/Pt(111) (Fig.~\ref{fig:S4}).}
        \label{fig:S5}
\end{figure}

We also measured the dependence of the FF-SFG spectra on pulse energy. As shown in Fig.~\ref{fig:S5}, the signal intensity scales linearly with both IR (Fig.~\ref{fig:S5}(a)) and probe (Fig.~\ref{fig:S5}(b)) pulse energies, while the spectral characteristics, including peak position and linewidth, remain essentially unchanged. The maximum IR pulse energy used in these FF-SFG experiments (\SI{1}{nJ}) corresponds to a surface fluence of \SI{7.9e-3}{mJ.cm^{-2}}, which is more than four orders of magnitude lower than the estimated near-field fluence required to drive the 3--4 transition (\SI{100}{mJ.cm^{-2}}, see the main text), and is therefore far too weak to induce VLC. Consequently, VLC is absent under the present far-field irradiation conditions.

\newpage
\section{Comparison of forward- and backward-scattered TE-SFG spectra}
\begin{figure}[H]
    \centering
        \includegraphics[width=0.9\linewidth]{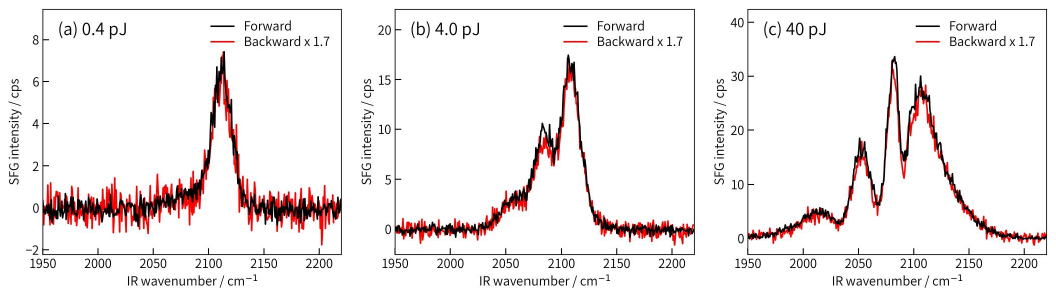}
        \caption{Comparison of forward- (black) and backward-scattered (red) TE-SFG spectra of CO/Pt(111) at
        three different IR pulse energies: (a) \SI{0.4}{pJ}, (b) \SI{4}{pJ}, and (c) \SI{40}{pJ}. The backward signal was multiplied by a factor of 1.7.}
        \label{fig:S6}
\end{figure}

In this section, we compare the forward- and backward-scattered TE-SFG signals, which were acquired simultaneously in our experiments (Fig.~\ref{fig:S3}). As shown in Fig.~\ref{fig:S6}, the two spectra exhibit nearly identical spectral shapes across three different IR pulse energies. This indicates that the detection geometry does not affect the main results of the TE-SFG measurements. Note that the forward-scattered TE-SFG signal is approximately \num{1.7} times stronger than the backward-scattered signal, likely due to the asymmetric geometry of the tip apex \cite{SI:AANM2021_Cui}. In the main text, we therefore present only the forward-scattered spectra, which provide a better signal-to-noise ratio.

\section{Estimating anharmonicity of CO stretching mode}
\begin{figure}[H]
        \centering
        \includegraphics[width=0.7\linewidth]{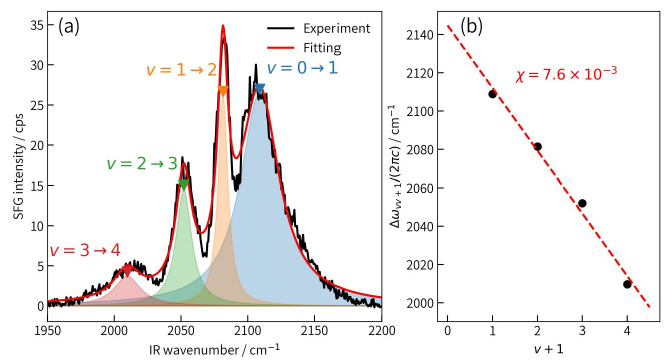}
        \caption{(a) TE-SFG spectrum of CO/Pt(111) obtained at an IR pulse energy of \SI{40}{pJ} (black curve), identical to that shown in Fig.~2(a) in the main text. The red curve represents the result of the multipeak fitting analysis, and the colored shaded areas indicate individual peak components. (b) Extracted vibrational transition energies plotted as a function of the vibrational quantum number. The red dashed line is a linear fit to the data.}
        \label{fig:S7}
\end{figure}

In this section, we quantitatively examine the anharmonicity of the atop CO stretching mode. As described in the main text, the vibrational progression observed in TE-SFG measurements for CO/Pt(111) exhibits nearly uniform spacing between adjacent peaks (Fig.~2(a)). This behavior indicates that the vibrational potential of the atop CO stretching mode can be well approximated by a Morse potential, in which the transition frequency for the $v \to v+1$ transition, $\Delta\omega_{v+1,v}$, decreases linearly with the vibrational quantum number $v$:
\begin{align}
    \Delta\omega_{v+1,v}&=\omega_{\text{e}}\left(1-2\chi(v+1)\right), \label{eq:anharmonicity}
\end{align}
where $\omega_{\text{e}}$ is the equilibrium frequency and $\chi$ is the anharmonicity constant \cite{SI:PR1929_Morse}. To determine $\omega_{\text{e}}$ and $\chi$, the vibrational transition energies were extracted by multipeak fitting of the TE-SFG spectrum (Fig.~\ref{fig:S7}(a)) and plotted as a function of the vibrational quantum number $v+1$ (Fig.~\ref{fig:S7}(b)). Fitting the data using Eq.~\eqref{eq:anharmonicity} (red line in Fig.~\ref{fig:S7}(b)) yields $\omega_{\text{e}}/(2\pi c)=\SI{2144.8}{cm^{-1}}$ and $\chi=\num{7.6e-3}$, in good agreement with previously reported values \cite{SI:JMS1975_Mantz,SI:JPCC2025_Omiya,SI:SS2005_Zhang,SI:CPL2000_Hess} (Table~\ref{tab:anharm_params_vs_refs}).

The numerical analysis based on the optical Bloch equations (Eq.~(1) in the main text) and the calculation of SFG spectra require the matrix element of the vibrational normal coordinate $\xi$ (i.e., $\bra{v+1}\xi\ket{v}$), which enters key physical quantities such as the transition dipole moments $\mu_{v+1,v}$, population relaxation rates $\Gamma_v$, and Raman polarizabilities $\alpha_{v+1,v}$. This matrix element was calculated using the following relation:
\begin{align}
    \bra{v+1}\xi\ket{v}&=c_{\text{corr}}(v;\chi)\frac{\sqrt{v+1}}{\sqrt{2}}, \label{eq:v1xiv} \\
    c_{\text{corr}}(v;\chi)&\equiv\sqrt{\frac{(1-(2v+1)\chi)(1-(2v+3)\chi)}
                                        {(1-2(v+1)\chi)^2(1-(v+1)\chi)}}. \label{eq:correction_factor}
\end{align}
The term $\sqrt{v+1}/\sqrt{2}$ corresponds to the harmonic approximation, whereas extension to the Morse potential introduces an anharmonic correction factor $c_{\text{corr}}(v;\chi)$ (Eq.~\ref{eq:v1xiv}) \cite{SI:JCP1983_Vasan,SI:PNAS2004_Ventalon}. The numerical values of $c_{\text{corr}}(v;\chi)$, evaluated by substituting $\chi=\num{7.6e-3}$ into Eq.~\eqref{eq:correction_factor}, are summarized in Table~\ref{tab:morse_anharmonic_correction_factors}.
\begin{table}[H]
  \centering
  \caption{Anharmonic parameters of the CO stretching mode. Estimated values reported in several previous studies are also shown.}
  \label{tab:anharm_params_vs_refs}
\begin{tabular}{cS[table-format=4.1]S[table-format=1.1e-1]}
\hline
\hline
     system & {$\omega_{\text{e}}/(2\pi c)$ / \si{cm^{-1}}} & {$\chi$}  \\
    \hline
    this study & 2144.8 & 7.6e-3 \\
    gas CO\cite{SI:JMS1975_Mantz} & 2169.8 & 6.1e-3 \\
    CO/Cu(110)\cite{SI:JPCC2025_Omiya} & 2114.2 & 6.1e-3 \\
    CO/Ir(111)\cite{SI:SS2005_Zhang} & 2067.8 & 6.5e-3 \\
    CO/Ru(001)\cite{SI:CPL2000_Hess} & 2015.6 & 6.7e-3 \\
\hline
\hline
\end{tabular}
\end{table}
\begin{table}[H]
  \centering
  \caption{Anharmonic correction factors $c_{\text{corr}}(v; \chi)$ based on Morse potential with $\chi=\num{7.6e-3}$.}
  \label{tab:morse_anharmonic_correction_factors}
  \begin{tabular}{SS}
  \hline
  \hline
  {$v$} & {$c_{\text{corr}}(v; \chi)$} \\
  \hline
    0 & 1.004 \\
    1 & 1.008 \\
    2 & 1.012 \\
    3 & 1.016 \\
    4 & 1.020 \\
    5 & 1.024 \\
    6 & 1.028 \\
\hline
\hline
\end{tabular}
\end{table}

\newpage
\section{Broadening of 0--1 band induced by intense IR field}
\begin{figure}[H]
    \centering
        \includegraphics[width=0.9\linewidth]{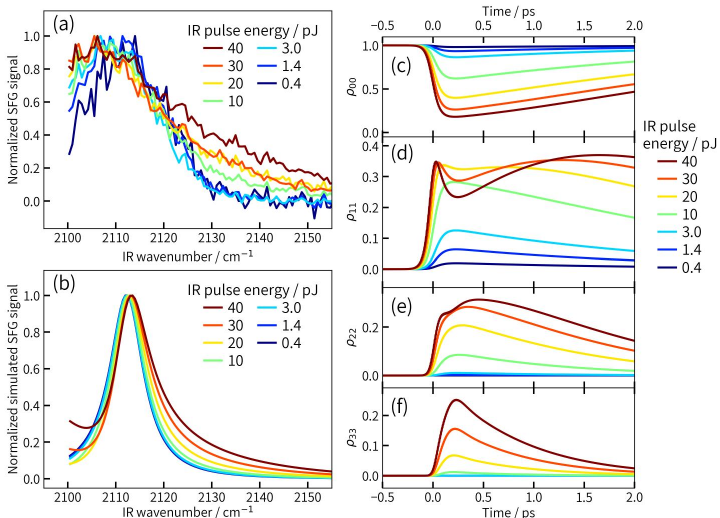}
        \caption{(a) Experimental and (b) simulated TE-SFG spectra of CO/Pt(111) at various IR pulse energies. The spectra are normalized to the maximum intensity of the 0--1 band.
        Time evolution of (c) $\rho_{00}$, (d) $\rho_{11}$, (e) $\rho_{22}$, and (f) $\rho_{33}$,
        simulated by numerically solving the multilevel optical Bloch equations (Eq.~(1) in the main text) at various IR pulse energies.}
        \label{fig:S8}
\end{figure}

As shown in Fig.~\ref{fig:S8}(a), the 0--1 band in the experimental TE-SFG spectra broadens with increasing IR pulse energy. This broadening is also captured in the numerically simulated SFG spectra (Fig.~\ref{fig:S8}(b)). To investigate its origin, we analyzed the simulated time evolution of the populations of the four lowest vibrational states (Fig.~\ref{fig:S8}(c--f)). In the time trace of the $\rho_{11}$ population, a decrease around $t=\SI{0}{ps}$ becomes evident for IR pulse energies above \SI{20}{pJ} (Fig.~\ref{fig:S8}(d)). This reduction in $\rho_{11}$ coincides with the buildup of the $\rho_{22}$ and $\rho_{33}$ populations (Fig.~\ref{fig:S8}(e), (f)). Therefore, an apparent shortening of the $\rho_{11}$ lifetime, caused by excitation to higher-lying vibrational levels, contributes to the broadening of the 0--1 band.

Comparing the linewidths of individual peaks in the vibrational progression (Fig.~\ref{fig:S7}(a)), the 1--2 band is clearly narrower than the 0--1 band, while the linewidth gradually increases for higher-order transitions. The broadening observed for higher vibrational transitions from 1--2 to 3--4 can be explained by the increase in the vibrational relaxation rate at higher vibrational levels, described by $\Gamma_{v+1}=\Gamma_1(\bra{v+1}\xi\ket{v})^2/(\bra{1}\xi\ket{0})^2$. Within this framework, however, the 0--1 band is expected to exhibit the narrowest linewidth, contrary to the experimental observations. Furthermore, such broadening limited to the 0--1 band has not been reported in previous far-field VLC studies of CO on metal surfaces \cite{SI:JPCC2025_Omiya,SI:SS2005_Zhang,SI:CPL2000_Hess}. These observations suggest the presence of additional nanogap-specific mechanisms that selectively shorten the lifetime of the 0--1 vibrational coherence, leading to its anomalous broadening. 
As noted in the main text, the broadened 0--1 band may arise from combined physical effects that are not included in our model, including intermolecular dipole--dipole coupling and the resulting delocalized vibrational excitations in the densely packed CO adlayer \cite{SI:Hamm_Zanni_2011}, instantaneous frequency shifts and associated pure dephasing induced by anharmonic coupling between the CO stretching mode and low-frequency modes \cite{SI:JCP2012_Inoue,SI:PRL2016_Inoue}, and the spatial nonuniformity of the tip-enhanced near field.
Although elucidating such microscopic light--molecule dynamics in the nanogap is beyond the scope of this work, further investigation of these dynamics remains an important direction for future research.

\newpage
\section{Delay dependence of TE-SFG spectra}
\begin{figure}[H]
    \centering
        \includegraphics[width=0.85\linewidth]{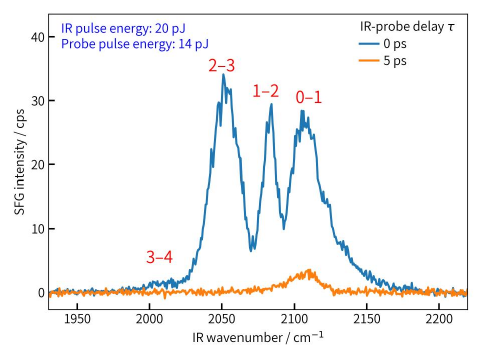}
        \caption{Experimental TE-SFG spectra measured at IR--probe delays of $\tau=0$ and \SI{5}{ps}, using IR and probe pulse energies of \SI{20}{pJ} and \SI{14}{pJ}, respectively. At $\tau=0$, the 0--1, 1--2, 2--3, and 3--4 bands form a distinct multipeak structure. At $\tau=\SI{5}{ps}$, the hot-band signals have almost completely disappeared, leaving only a weak 0--1-band signal. Because \SI{5}{ps} is far shorter than the \SI{20}{ns} interval between successive laser pulses, the disappearance rules out population accumulation over successive pulses and supports vibrational ladder climbing driven by a single IR pulse as an intra-pulse excitation process.}
        \label{fig:S9}
\end{figure}

To directly examine whether the vibrational ladder climbing (VLC) reported in the main text is generated and probed within a single infrared (IR) pulse or instead requires accumulation over successive pulses, we measured TE-SFG spectra of CO/Pt(111) as a function of the delay $\tau$ between the IR pump and the upconversion probe pulse. Figure~\ref{fig:S9} compares spectra acquired at IR--probe delays of $\tau=0$ and \SI{5}{ps}, using IR and probe pulse energies of \SI{20}{pJ} and \SI{14}{pJ}, respectively. At $\tau=0$, the 0--1, 1--2, 2--3, and 3--4 bands form a distinct multipeak structure, consistent with the spectra shown in Fig.~2(a) of the main text. At $\tau=\SI{5}{ps}$, the hot-band signals have almost completely disappeared, leaving only a weak 0--1-band signal.

The hot-band signals almost completely disappear by $\tau=\SI{5}{ps}$, which is far shorter than the \SI{20}{ns} interval between successive laser pulses at the \SI{50}{MHz} repetition rate. This result rules out population accumulation over successive pulses and provides experimental support for vibrational ladder climbing driven by a single IR pulse, namely, an intra-pulse excitation process. Because several sequential transitions are driven during the few-hundred-femtosecond interval in which the enhanced IR field is present, the observation is consistent with a coherent light--vibration interaction.

\section{Simulated spectra with broader-bandwidth IR pulses}

\begin{figure}[H]
    \centering
        \includegraphics[width=0.9\linewidth]{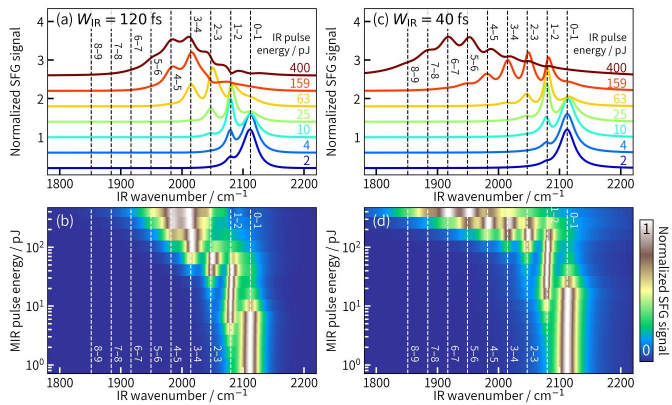}
        \caption{Simulated TE-SFG spectra at various IR pulse energies. The only modification to the model used in the main-text simulations is an expansion from the seven-level basis ($v=0$ through $v=6$) to a 12-level basis ($v=0$ through $v=11$) in all panels, which allows population transfer to higher vibrational levels to be examined. All other model parameters are unchanged. The IR pulses are assumed to be Fourier-transform-limited $\mathrm{sech}^{2}$ pulses, and their durations are given as the FWHM of the intensity envelope $|E(t)|^{2}$. (a, b) Results for the \SI{120}{fs} IR pulse used in the experiments (bandwidth $\mathord{\sim}\SI{90}{cm^{-1}}$ FWHM) reproduce the sequential emergence of hot bands up to the 3--4 transition as described in the main text. (c, d) Results for a broader-bandwidth \SI{40}{fs} IR pulse ($\mathord{\sim}\SI{260}{cm^{-1}}$ FWHM) predict that the 7--8 transition would become observable. This comparison indicates that the experimentally accessible level is limited by the IR excitation bandwidth. Panels (a) and (c) show vertically offset spectra, while panels (b) and (d) show the corresponding 2D maps as functions of wavenumber and IR pulse energy. All spectra are normalized independently at each IR pulse energy.}
        \label{fig:S10}
\end{figure}

\newpage

\makeatletter
\def\NAT@bibsetnum#1{%
  \setlength{\topsep}{\z@}%
  \NATx@bibsetnum{\ref{LastBibItemSI}}%
}
\renewenvironment{thebibliography}[1]{%
  \NAT@thebibliography{#1}%
  \let\@TBN@opr\present@bibnote
  \@FMN@list
}{%
  \auto@bib@innerbib
  \edef\@currentlabel{\arabic{NAT@ctr}}%
  \label{LastBibItemSI}%
  \endNAT@thebibliography
  \aftergroup\auto@bib@empty
}
\makeatother

\makeatother

\end{document}